\documentstyle[12pt]{article}

\def\beq{\begin{equation}}
\def\eeq{\end{equation}}
\def\bea{\begin{eqnarray}}
\def\eea{\end{eqnarray}}
\def\nn{\nonumber}
\def\sss{\scriptscriptstyle}

\def\bd{B_d^0}
\def\bdbar{{\overline{B_d^0}}}
\def\bs{B_s^0}
\def\bsbar{{\overline{B_s^0}}}
\def\barp{{\raise.35ex\hbox
{${\sss (}$}}---{\raise.35ex\hbox{${\sss )}$}}}
\def\bdbarp{\hbox{$B_d$\kern-1.4em\raise1.4ex\hbox{\barp}}}
\def\bsbarp{\hbox{$B_s$\kern-1.4em\raise1.4ex\hbox{\barp}}}
\def\ks{K_{\sss S}}

\def\roughly#1{\mathrel{\raise.3ex\hbox
{$#1$\kern-.75em\lower1ex\hbox{$\sim$}}}}

\def\gsim{\roughly>}

\def\Abar{\bar A}

\def\tnp{\theta_{\sss NP}}

\def\adirpm{{a_{\sss dir}^{+-}}}
\def\adir00{{a_{\sss dir}^{00}}}
\def\sineff{{\sin 2\,\alpha_{\sss eff}}}
\def\alphaeff{{\alpha_{\sss eff}}}
\def\alphaeffp{{\alpha_{\sss eff}^{(+)}}}
\def\alphaeffm{{\alpha_{\sss eff}^{(-)}}}
\def\alphaeffpm{{\alpha_{\sss eff}^{+-}}}
\def\alphaeffzero{{\alpha_{\sss eff}^{00}}}
\def\B00{B^{00}}
\def\Bp0{B^{+0}}
\def\Bpm{B^{+-}}
\def\dsp{\displaystyle}
\def\epjc#1#2#3{{\it Eur.\ Phys.\ J.}\ {\bf C#1} (19#2) #3} 
\def\ijmp#1#2#3{{\it Int.\ J.\ Mod.\ Phys.} {\bf A#1} (19#2) #3}

\def\npb#1#2#3{{\it Nucl.\ Phys.} {\bf B#1} (19#2) #3}
\def\plb#1#2#3{{\it Phys.\ Lett.} {\bf #1B} (19#2) #3}

\def\prd#1#2#3{{\it Phys.\ Rev.} {\bf D#1} (19#2) #3}
\def\newprd#1#2#3{{\it Phys.\ Rev.} {\bf D#1}: #3 (19#2)}
\def\newprdtwo#1#2#3{{\it Phys.\ Rev.} {\bf D#1}: #3 (20#2)}

\def\prl#1#2#3{{\it Phys.\ Rev.\ Lett.} {\bf #1} (19#2) #3}
\def\prltwo#1#2#3{{\it Phys.\ Rev.\ Lett.} {\bf #1} (20#2) #3}

\def\zpc#1#2#3{{\it Zeit.\ Phys.} {\bf C#1} (19#2) #3}

\newread\epsffilein % file to \read
\newif\ifepsffileok % continue looking for the bounding box?
\newif\ifepsfbbfound % success?
\newif\ifepsfverbose % report what you're making?
\newdimen\epsfxsize % horizontal size after scaling
\newdimen\epsfysize % vertical size after scaling
\newdimen\epsftsize % horizontal size before scaling
\newdimen\epsfrsize % vertical size before scaling
\newdimen\epsftmp % register for arithmetic manipulation
\newdimen\pspoints % conversion factor
\def\epsfbox#1{\global\def\epsfllx{72}\global\def\epsflly{72}%
 \global\def\epsfurx{540}\global\def\epsfury{720}%
 \def\lbracket{[}\def\testit{#1}\ifx\testit\lbracket
 \let\next=\epsfgetlitbb\else\let\next=\epsfnormal\fi\next{#1}}%
\def\epsfgetlitbb#1#2 #3 #4 #5]#6{\epsfgrab #2 #3 #4 #5 .\\%
 \epsfsetgraph{#6}}%
\def\epsfnormal#1{\epsfgetbb{#1}\epsfsetgraph{#1}}%
\def\epsfgetbb#1{%
\openin\epsffilein=#1
\ifeof\epsffilein\errmessage{I couldn't open #1, will ignore it}\else
 {\epsffileoktrue \chardef\other=12
 \def\do##1{\catcode`##1=\other}\dospecials \catcode`\ =10
 \loop
 \read\epsffilein to \epsffileline
 \ifeof\epsffilein\epsffileokfalse\else
 \expandafter\epsfaux\epsffileline:. \\%
 \fi
 \ifepsffileok\repeat
 \ifepsfbbfound\else
 \ifepsfverbose\message{No bounding box comment in #1; using defaults}\fi\fi
 }\closein\epsffilein\fi}%
\def\epsfclipstring{}% do we clip or not? If so,
\def\epsfsetgraph#1{%
 \epsfrsize=\epsfury\pspoints
 \advance\epsfrsize by-\epsflly\pspoints
 \epsftsize=\epsfurx\pspoints
 \advance\epsftsize by-\epsfllx\pspoints
 \epsfxsize\epsfsize\epsftsize\epsfrsize
 \ifnum\epsfxsize=0 \ifnum\epsfysize=0
 \epsfxsize=\epsftsize \epsfysize=\epsfrsize
 \epsfrsize=0pt
 \else\epsftmp=\epsftsize \divide\epsftmp\epsfrsize
 \epsfxsize=\epsfysize \multiply\epsfxsize\epsftmp
 \multiply\epsftmp\epsfrsize \advance\epsftsize-\epsftmp
 \epsftmp=\epsfysize
 \loop \advance\epsftsize\epsftsize \divide\epsftmp 2
 \ifnum\epsftmp>0
 \ifnum\epsftsize<\epsfrsize\else
 \advance\epsftsize-\epsfrsize \advance\epsfxsize\epsftmp \fi
 \repeat
 \epsfrsize=0pt
 \fi
 \else \ifnum\epsfysize=0
 \epsftmp=\epsfrsize \divide\epsftmp\epsftsize
 \epsfysize=\epsfxsize \multiply\epsfysize\epsftmp
 \multiply\epsftmp\epsftsize \advance\epsfrsize-\epsftmp
 \epsftmp=\epsfxsize
 \loop \advance\epsfrsize\epsfrsize \divide\epsftmp 2
 \ifnum\epsftmp>0
 \ifnum\epsfrsize<\epsftsize\else
 \advance\epsfrsize-\epsftsize \advance\epsfysize\epsftmp \fi
 \repeat
 \epsfrsize=0pt
 \else
 \epsfrsize=\epsfysize
 \fi
 \fi
 \ifepsfverbose\message{#1: width=\the\epsfxsize, height=\the\epsfysize}\fi
 \epsftmp=10\epsfxsize \divide\epsftmp\pspoints
 \vbox to\epsfysize{\vfil\hbox to\epsfxsize{%
 \ifnum\epsfrsize=0\relax
 \includegraphics{#1}%
 \else
 \epsfrsize=10\epsfysize \divide\epsfrsize\pspoints
 \includegraphics{#1}%
 \fi
 \hfil}}%
\global\epsfxsize=0pt\global\epsfysize=0pt}%
 {\catcode`\%=12 \global\let\epsfpercent=%\global\def\epsfbblit{%BoundingBox}}%
\long\def\epsfaux#1#2:#3\\{\ifx#1\epsfpercent
 \def\testit{#2}\ifx\testit\epsfbblit
 \epsfgrab #3 . . . \\%
 \epsffileokfalse
 \global\epsfbbfoundtrue
 \fi\else\ifx#1\par\else\epsffileokfalse\fi\fi}%
\def\epsfempty{}%
\def\epsfgrab #1 #2 #3 #4 #5\\{%
\global\def\epsfllx{#1}\ifx\epsfllx\epsfempty
 \epsfgrab #2 #3 #4 #5 .\\\else
 \global\def\epsflly{#2}%
 \global\def\epsfurx{#3}\global\def\epsfury{#4}\fi}%
\def\epsfsize#1#2{\epsfxsize}
\begin{document}

\setlength{\baselineskip}{20pt}

\begin{flushright}
UdeM-GPP-TH-00-77 \\
IMSc-99/05/18 \\
\end{flushright}

\begin{center}
\bigskip

{\Large \bf Searching for New Physics via CP Violation in $B\to\pi\pi$ } \\
\bigskip

David London $^{a,}$\footnote{london@lps.umontreal.ca},~~ 
Nita Sinha $^{b,}$\footnote{nita@imsc.ernet.in}~~
and Rahul Sinha $^{b,}$\footnote{sinha@imsc.ernet.in}
\end{center}

%\medskip

\begin{flushleft}
~~~~~~~~~~~$a$: {\it Laboratoire Ren\'e J.-A. L\'evesque, 
Universit\'e de Montr\'eal,}\\
~~~~~~~~~~~~~~~{\it C.P. 6128, succ. centre-ville, Montr\'eal, QC,
Canada H3C 3J7}\\
~~~~~~~~~~~$b$: {\it Institute of Mathematical Sciences, Taramani,
 Chennai 600113, India}
\end{flushleft}

\begin{center}
 
\bigskip (\today)

\bigskip 

{\bf Abstract}

\end{center}

\begin{quote} 
  It is well known that one can use $B\to\pi\pi$ decays to probe the
  CP-violating phase $\alpha$. In this paper we show that these same
  decays can be used to search for new physics. This is done by
  comparing two weak phases which are equal in the standard model: the
  phase of the $t$-quark contribution to the $b\to d$ penguin
  amplitude, and the phase of $\bd$--$\bdbar$ mixing. In order to make
  such a comparison, we require one piece of theoretical input, which
  we take to be a prediction for $|P/T|$, the relative size of the
  penguin and tree contributions to $\bd\to\pi^+\pi^-$. If independent
  knowledge of $\alpha$ is available, the decay $\bd(t)\to\pi^+\pi^-$
  alone can be used to search for new physics.  If a full isospin
  analysis can be done, then new physics can be found solely through
  measurements of $B\to\pi\pi$ decays. The most promising scenario
  occurs when the isospin analysis can be combined with independent
  knowledge of $\alpha$. In all cases, the prospects for detecting new
  physics in $B\to\pi\pi$ decays can be greatly improved with the help
  of additional measurements which will remove discrete ambiguities.
\end{quote}
\newpage

\section{Introduction}

%\noindent {\bf Sat., Apr. 1, 2000}

Within the standard model (SM), CP violation is due to the presence of
a nonzero complex phase in the Cabibbo-Kobayashi-Maskawa (CKM) quark
mixing matrix. This phase information can be elegantly displayed using
the unitarity triangle \cite{PDG98}, in which the interior
(CP-violating) angles are called $\alpha$, $\beta$ and $\gamma$. In
the near future, these CP angles will be extracted from the
measurements of rate asymmetries in $B$ decays \cite{CPreview}. As
usual, the hope is that these measurements will reveal the presence of
physics beyond the SM.

The canonical decay modes which will be used to measure the CP angles
are $\bd(t) \to \pi^+ \pi^-$ ($\alpha$), $\bd(t) \to J/\Psi \ks$
($\beta$) and $B^\pm \to D K^\pm$ ($\gamma$) \cite{BDK}. Assuming that
each decay is dominated by a single amplitude, the corresponding CP
angle can be extracted with no hadronic uncertainties. Unfortunately,
this assumption does not hold for the decay $\bd \to \pi^+ \pi^-$: in
addition to the tree contribution $T$, there is also a penguin
contribution $P$ which may be sizeable \cite{penguins}. Nevertheless,
it is still possible to obtain $\alpha$ without hadronic uncertainties
if a $B \to \pi\pi$ isospin analysis can be performed \cite{isospin}.

If new physics is present, it will contribute principally at
loop-level, affecting $B_q^0$--${\overline{B_q^0}}$ mixing ($q=d,s$)
\cite{NPBmixing} and/or the $b\to q$ flavour-changing neutral current
(FCNC) penguin amplitudes \cite{GroWorah}. There are a variety of ways
of detecting this new physics. For example, if there is an
inconsistency between the unitarity triangle as constructed from
measurements of the angles and that constructed from independent
measurements of the sides, this will be a signal of new physics.
However, there are two potential difficulties with this. First,
measurements of the sides of the triangle require theoretical input
regarding certain hadronic quantities. Depending on the size of the
discrepancy, one might question the precision of the theoretical
numbers. Second, there are discrete ambiguities in extracting the
angles, and it may be necessary to resolve these ambiguities in order
to be certain that a discrepancy is in fact present \cite{disamb}.

A more direct way of looking for new physics is to consider two CP
asymmetries which in the SM are supposed to probe the same CP
angle. For example, the angle $\gamma$ can be measured via $B^\pm \to
D K^\pm$ \cite{BDK} or $\bs(t) \to D_s^\pm K^\mp$ \cite{BsDsK}. If the
measured values of $\gamma$ in these two modes disagree, this is a
clear sign of new physics. Similarly, the angle $\beta$ can be
measured via $\bd(t) \to J/\Psi \ks$ or $\bd(t) \to \phi \ks$
\cite{LonSoni}. Another example, similar in spirit to these, is the
decay $\bs(t) \to J/\Psi \phi$. Within the SM, the CP asymmetry in
this decay is expected to vanish, so that a nonzero value would
indicate the presence of new physics.

In all of these examples, the new physics affects the $b\to s$ FCNC,
either through $\bs$-$\bsbar$ mixing or the $b \to s$ penguin
amplitude. One might then wonder whether it is possible to detect new
physics in this way if it affects only the $b \to d$ FCNC. For
example, in the Wolfenstein parametrization \cite{Wolfenstein}, the
weak phase of $\bd$-$\bdbar$ mixing and of the $t$-quark contribution
to the $b\to d$ penguin are both equal to $-\beta$ in the SM. However,
in the presence of new physics, these phases could be different.
Therefore, if one could measure these two phases and find a
discrepancy, this would be a clear signal of new physics.

Unfortunately, in a recent paper \cite{LSS}, we showed that this is
not possible. In the SM, the largest contribution to the $b \to d$
penguin comes from an internal $t$-quark, and is proportional to
$V_{tb}^* V_{td}$. However, the contributions of an internal $u$-quark
($V_{ub}^* V_{ud}$) and an internal $c$-quark ($V_{cb}^* V_{cd}$) are
not negligible \cite{ucquark}. It is therefore impossible to isolate
any single contribution: using the unitarity of the CKM matrix,
\beq
V_{ub}^* V_{ud} + V_{cb}^* V_{cd} + V_{tb}^* V_{td} = 0 ~,
\label{unitarity}
\eeq
it is always possible to write one amplitude in terms of the other
two. And because one cannot isolate the $t$-quark amplitude, one
cannot cleanly measure its phase. In Ref.~\cite{LSS}, we refer to this
as the ``CKM ambiguity.'' However, we also note that the CKM ambiguity
can be resolved if one makes an assumption regarding the hadronic
parameters involved in the $b \to d$ FCNC amplitude.

In this paper, we apply this idea to $\bd(t) \to \pi^+\pi^-$. As
mentioned earlier, this decay receives contributions from both a
tree-level amplitude $T$ and a $b\to d$ penguin amplitude $P$. The
isospin analysis essentially allows one to remove this penguin
``pollution'' and hence obtain a clean measurement of $\alpha$. Of
course, as argued above, there is not enough information to extract
the phase of the $t$-quark contribution to the $b\to d$ penguin.
However, if we make an assumption about the relative size of $P$ and
$T$, this provides us with the additional piece of information
necessary to test for the presence of new physics. As we will show, in
principle this method does indeed work: given an assumption about
$|P/T|$, the isospin analysis can be used not only to obtain $\alpha$
cleanly, but also to see if new physics is present.

In fact, the isospin analysis is not even necessary. In the absence of
new physics, the ratio $|P/T|$ depends only on $\alpha$ and the
quantities measured in $\bd(t) \to \pi^+\pi^-$ \cite{Charles}. Thus,
if independent information about $\alpha$ is available, the
measurement of $\bd(t) \to \pi^+\pi^-$ alone will suffice to obtain
$|P/T|$. If new physics is present, and affects the magnitude of $P$,
then obviously the extracted value of $|P/T|$ will differ from its SM
value. However, a more interesting scenario is if the only effect of
new physics is to produce a discrepancy between the weak phase of
$\bd$-$\bdbar$ mixing and that of the $t$-quark contribution to the
$b\to d$ penguin. What is perhaps not obvious, but is in fact true, as
we will show, is that even in this case, the extracted value of
$|P/T|$ will still differ from that which one would have obtained in
the absence of new physics. Therefore, given a prediction for $|P/T|$
and some knowledge of $\alpha$ (either from independent measurements
or via an isospin analysis), the measurement of $\bd(t) \to
\pi^+\pi^-$ can be used to search for new physics in the $b\to d$
FCNC.

Not surprisingly, however, there are some potential problems which
must be taken into consideration. Most importantly, there are discrete
ambiguities in extracting some of the phases necessary for the
analysis. Their presence may make the discovery of new physics
difficult, particularly since there will be errors associated with
both the experimental measurements and theoretical predictions. In
order to remove discrete ambiguities, it is necessary to be able to
measure the same quantities in a variety of ways. For example, the
search for new physics will be facilitated if we have independent
information about $\alpha$ {\it and} are able to perform an isospin
analysis. However, it may happen that, due to small branching ratios
or poor detection efficiencies, one cannot measure the rates for the
decays $\bd/\bdbar \to \pi^0\pi^0$ and/or $B^+ \to \pi^+ \pi^0$.
Instead, only upper limits can be obtained, so that a full isospin
analysis cannot be performed. In this case, one has to examine the
extent to which partial knowledge of these quantities helps in
detecting the presence of new physics.

In this paper, we discuss all of these issues. We begin in Sec.~2 with
a review of the $B\to \pi\pi$ isospin analysis. Here we show how new
physics affects the extraction of $|P/T|$, and present the SM
expectations for the magnitude of this ratio. We also discuss the
potential difficulties (discrete ambiguities, incomplete isospin
analysis) in looking for new physics by combining the isospin analysis
with a theoretical prediction for $|P/T|$. In the following two
sections, we take $|P/T|$ to lie within a particular range of values,
and examine the prospects for detecting the presence of new physics in
$\bd(t) \to \pi^+\pi^-$. In Sec.~3 it is assumed that only $\bd(t) \to
\pi^+\pi^-$ has been measured. Here we also require independent
information about $2\alpha$. In this scenario, it is possible to
detect the presence of new physics, but there are complications due to
discrete ambiguities. The prospects for detecting new physics can be
significantly improved with the help of other, independent
measurements which can be used to remove these discrete ambiguities.
In Sec.~4 we examine the effect of combining the measurement of
$\bd(t) \to \pi^+\pi^-$ with an isospin analysis. Surprisingly, even
if no information is available regarding $B^+\to\pi^+\pi^0$ and
$\bd/\bdbar \to \pi^0\pi^0$ decays, the isospin symmetry nevertheless
reduces the discrete ambiguities found in Sec.~3 by a factor of two.
Of course, the situation is improved if we do have information about
$B^+\to\pi^+\pi^0$ and $\bd/\bdbar \to \pi^0\pi^0$. This can take one
of two forms.  Either a full isospin analysis is possible, which
involves measuring the decays $B^+\to\pi^+\pi^0$ and $\bd/\bdbar \to
\pi^0\pi^0$, or we have only limits on the quantities involved in
these decays. In either case, one can indeed detect the presence of
new physics, but once again discrete ambiguities complicate matters.
The situation can be greatly improved if one assumes, as is likely to
be the case in practice, that independent information about $2\alpha$
is available.  We conclude in Sec.~5.

\section{Theoretical Framework}

\subsection{Isospin Analysis}

We begin with a review of the $B\to \pi\pi$ isospin analysis. In the
SM, in the Wolfenstein parametrization, the weak phase of the
$\bd$--$\bdbar$ mixing amplitude is $e^{-2 i \beta}$. When considering
$B$ decays, it is useful to remove this mixing phase by redefining the
decay amplitudes as follows:
\beq 
A^f \equiv e^{i \beta} Amp(\bd \to f) ~~,~~~~ \Abar^f \equiv e^{-i\beta}
Amp({\bar \bd} \to f) ~.
\eeq
Then the time-dependent decay rate for a $\bd(t)$ to decay into a
final state $f$ takes the form
\beq
\Gamma(\bd(t) \to f) = e^{-\Gamma t} \left[ {|A^f|^2 + |\Abar^f|^2 \over 2}
+ {|A^f|^2 - |\Abar^f|^2 \over 2} \cos (\Delta M t) 
- {\rm Im} \left({A^f}^* \Abar^f\right) \sin (\Delta M t) \right],
\label{timedep}
\eeq
where $\bd(t)$ is a $B$-meson which at $t=0$ was a $\bd$. 

In general, the decay $\bd \to \pi \pi$ receives contributions from a
tree-level amplitude and a $b\to d$ penguin amplitude. Using the
unitarity of the CKM matrix to eliminate the $V_{cb}^* V_{cd}$ piece
of the penguin diagram, we can write
\beq
A(\bd \to \pi^+ \pi^-)  \equiv  A^{+-} 
 =   \sqrt{2} \left[\,T e^{i \delta} e^{-i \alpha} 
  + P e^{i \delta_{\sss P}} e^{-i\tnp}\right],
\label{+-amp}
\eeq
where the $T e^{i \delta}$ term includes the $u$-quark piece
of the penguin amplitude, and $P e^{i \delta_{\sss P}}$
contains the remaining contributions to the penguin amplitudes.  The
$\delta$'s are strong phases and the electroweak penguin contribution
has been ignored \cite{EWPs}. In Eq.~(\ref{+-amp}), we have allowed
for the possibility of new physics affecting the $b\to d$ FCNC by
including the new-physics phase $\tnp$. This phase will be nonzero if
the $\bd$--$\bdbar$ mixing amplitude and the $b\to d$ penguin
amplitude are affected by the new physics in different ways. (Note
that it is also possible for new physics to affect the magnitudes of
$T$ and $P$. This possibility is implicitly included in our
method.) The corresponding $\Abar^{+-}$ amplitude is obtained from the
$A^{+-}$ amplitude in Eq.~(\ref{+-amp}) by changing the signs of the
weak phases $\alpha$ and $\tnp$.

If the penguin contributions to $\bd\to\pi^+\pi^-$ are negligible,
then the measurement of the time-dependent rate for this decay allows
one to obtain the CP angle $\alpha$. From Eq.~(\ref{timedep}), the
coefficient of the $\sin(\Delta M t)$ term probes the relative phase
of the $A^{+-}$ and $\Abar^{+-}$ amplitudes. And, from Eq.~\ref{+-amp}
we see that this relative phase is $2\alpha$ if $P \sim 0$. On
the other hand, if $P$ is not negligible, then $\alpha$ cannot be
cleanly extracted from this measurement, since the relative phase of
$A^{+-}$ and $\Abar^{+-}$ is then a complicated function of $\alpha$
and the other parameters.

Under such circumstances, an isospin analysis can be used to cleanly
extract $\alpha$. The amplitude for $\bd\to\pi^+\pi^-$ is related by
isospin to the amplitudes for $\bd\to\pi^0\pi^0$ ($A^{00}$) and
$B^+\to\pi^+\pi^0$ ($A^{+0}$):
\beq
A^{+0} = {1\over \sqrt{2}} \, A^{+-} + A^{00} ~.
\label{Atriangle}
\eeq
Thus, if we write
\bea
\label{00amp}
A^{00} & = & T^{00} e^{i \delta^{00}} e^{-i \alpha} 
  + P^{00} e^{i \delta_{\sss P}^{00}} e^{-i\tnp} ~, \\
\label{+0amp}
A^{+0} & = & T^{+0} e^{i \delta^{+0}} e^{-i \alpha} ~,
\eea
the isospin relation [Eq.~(\ref{Atriangle})] implies
\bea
T^{+0} e^{i \delta^{+0}} &=& T e^{i \delta} +
T^{00} e^{i \delta^{00}}, ~\nn\\
P^{00} e^{i \delta_{\sss P}^{00}} &=& - P e^{i \delta_{\sss P}} ~. 
\label{isospin}
\eea
The $\Abar$ amplitudes obey similar isospin relations.

In order to obtain $\alpha$, we note that the magnitudes of the six
amplitudes $|A^{+-}|$, $|A^{00}|$, $|A^{+0}|$, $|\Abar^{+-}|$,
$|\Abar^{00}|$ and $|\Abar^{-0}|$ can be measured experimentally. We
can therefore construct the isospin triangles involving the $A$ and
$\Abar$ amplitudes. Furthermore, as noted above, the relative phase
between the $A^{+-}$ and $\Abar^{+-}$ amplitudes can be measured in
$\bd(t)\to\pi^+\pi^-$. This then fixes the relative orientations of
the $A$- and $\Abar$-triangles. However, the key point here is that
this also fixes the relative orientations of the $A^{+0}$ and
$\Abar^{-0}$ amplitudes. Since the relative phase of these two
amplitudes is just $2\alpha$ [see Eq.~(\ref{+0amp})], this shows that
the isospin analysis allows one to remove the penguin pollution and
cleanly extract $\alpha$.

Explicitly, $\alpha$ is found as follows. First, we define the
relative phase between the $A^{+-}$ and ${\bar A}^{+-}$ amplitudes to
be $2\alphaeffpm$.  Second, the construction of the $A$-triangle
allows one to measure $\Phi$, the angle between the $A^{+0}$ and
$A^{+-}$ amplitude. Similarly, the $\Abar$-triangle can be used to
obtain ${\bar\Phi}$, the angle between ${\bar A}^{-0}$ and ${\bar
A}^{+-}$. $\Phi$ and ${\bar\Phi}$ are defined via
\begin{eqnarray}
\cos\Phi & = & \frac{ ({1\over 2} |A^{+-}|^2 + |A^{+0}|^2 - |A^{00}|^2)} 
{\sqrt{2}|A^{+-}||A^{+0}|} ~, \nn \\
\cos\bar{\Phi} & = & \frac{ ( {1\over 2} |\bar{A}^{+-}|^2 + |\bar{A}^{-0}|^2 
- |\bar{A}^{00}|^2)} {\sqrt{2}|\bar{A}^{+-}||\bar{A}^{-0}|} ~.
\label{cosdefs}
\end{eqnarray}
Given that $2\alpha$ is the relative phase between $A^{+0}$ and
$\Abar^{-0}$, the angle $\alpha$ is then determined by
$2\alpha=2\alphaeffpm+\bar{\Phi}-\Phi$.

Finally, it is useful to examine which measurements are really needed
in order to carry out the isospin analysis. This analysis involves six
amplitudes: $A^{+-}$, $A^{00}$, $A^{+0}$, $\Abar^{+-}$, $\Abar^{00}$
and $\Abar^{-0}$. Experimentally, at best one can measure the
magnitudes and relative phases of these six amplitudes, giving 11
measurements. However, due to the (complex) $A$ and $\Abar$ isospin
triangle relations, four of the measurements are not independent.
Furthermore, we have $|A^{+0}| = |\Abar^{-0}|$.  Thus, of the 11
measurements, only six are independent. Three of these come from
measurements of $\bd(t) \to \pi^+\pi^-$:
\bea 
B^{+-} &\equiv& {1\over 2} \left( |A^{+-}|^2+|{\bar A^{+-}}|^2 \right) ~,~\nn\\
\adirpm &\equiv& {{|A^{+-}|^2-|{\bar A^{+-}}|^2} \over
{|A^{+-}|^2+|{\bar A^{+-}}|^2} }~,~\nn\\
2\alphaeffpm &\equiv&
Arg\left({A^{+-}}^* \Abar^{+-}\right) ~.
\eea
Two more can be obtained from measurements of $\bd \to \pi^0\pi^0$ and
$\bdbar\to \pi^0\pi^0$. They are $B^{00}$ and $\adir00$, defined
analogously to the above expressions for $B^{+-}$ and $\adirpm$. The
sixth measurement is taken to be the branching ratio for $B^+ \to
\pi^+ \pi^0$, $B^{+0} \equiv |A^{+0}|^2$. (Note that, since $|A^{+0}|
= |\Abar^{-0}|$, $B^{+0}$ is equal to $B^{-0}$, the branching ratio
for $B^- \to \pi^- \pi^0$.) In principle, the measurement of
$\bd(t)\to\pi^0\pi^0$ would also allow one to measure $\sin
(2\alphaeffzero)$. In practice, however, this is unlikely to be
feasible. And in any case, since there are only six independent
measurements, $\sin (2\alphaeffzero)$ can always be expressed in terms
of the other measurements. We will thus refer to $\alphaeffpm$ as
$\alphaeff$ from now on.

In terms of measurable quantities, the quantities $\cos\Phi$ and
$\cos\bar{\Phi}$ defined in Eq.~(\ref{cosdefs}) can be expressed as
\begin{eqnarray}
\cos\Phi & = & { {1\over 2} B^{+-} (1 + \adirpm) + B^{+0} - B^{00} (1 + \adir00)
\over \sqrt{2} \sqrt{B^{+-} (1 + \adirpm)} \sqrt{B^{+0}} } ~, \nn \\
\cos\bar{\Phi} & = & { {1\over 2} B^{+-} (1 - \adirpm) + B^{+0} - B^{00} (1 - \adir00)
\over \sqrt{2} \sqrt{B^{+-} (1 - \adirpm)} \sqrt{B^{+0}} } ~.
\end{eqnarray}
Note that these quantities depend only on ratios of branching
ratios. Thus, the isospin analysis can be carried out with knowledge
of only five of the six independent quantities. These are:
$B^{00}/B^{+-}$, $B^{+0}/B^{+-}$, $\adirpm$, $\adir00$ and
$2\alphaeff$. (Of course, in practice, all six independent
measurements will be made.)

\subsection{New Physics}

The theoretical expressions for the amplitudes [Eqs.~(\ref{+-amp}),
(\ref{00amp}) and (\ref{+0amp})] contain a total of seven physical
parameters: $\alpha$, $\tnp$, $T$, $T^{00}$, $P$, $\Delta \equiv
\delta - \delta_{\sss P}$ and $\Delta^{00} \equiv \delta^{00} -
\delta_{\sss P}$. With only six experimental measurements, it is
obvious that one cannot solve for all these parameters (this was to be
expected, given the CKM ambiguity \cite{LSS}). However, it is useful
to express some of these parameters in terms of the measurable
quantities and the angles $\alpha$ and $\tnp$. In particular, we have
\bea
P^2 & = & { B^{+-} \over 4 \sin^2 (\alpha-\tnp)}
\left[1-y\cos(2 \alpha - 2 \alphaeff)\right] ~, 
\label{P+-} \\ 
T^2 & = & { B^{+-} \over 4 \sin^2 (\alpha-\tnp)}
\left[1-y \cos(2 \tnp - 2 \alphaeff)\right] ~,
\label{T+-}
\eea
where we have defined $y \equiv\sqrt{1-(\adirpm)^2}$. (Expressions
similar to these, for the case $\tnp=0$, were first derived in
Ref.~\cite{Charles}.)

The ratio of the magnitudes of the penguin and tree amplitudes for the
$\bd \to \pi^+ \pi^-$ mode then has the following simple functional
form in terms of $\tnp$:
\beq
r^2 \equiv \frac {P^2}{T^2} =
{1-y \cos(2 \alpha - 2 \alphaeff) \over 1-y \cos(2 \tnp - 2 \alphaeff)} ~.
\label{r}
\eeq 
{}From this expression, we can see that, given measurements of
$\alpha$, $\adirpm$ and $\alphaeff$, and given a theoretical
prediction for $r$, one can obtain $\tnp$. (Note that a full isospin
analysis is not necessary here. If $\alpha$ can be obtained from
measurements outside the $B\to\pi\pi$ system, then the measurement of
$\bd(t)\to\pi^+\pi^-$ is sufficient to obtain $r^2$.) Obviously, if it
is found that $\tnp \ne 0$, this will indicate the presence of new
physics. (Note that if, in reality, $\tnp=0$ but new physics has
affected the magnitudes of $P$ and $T$, this may still show
up as an effective nonzero $\tnp$. But since we are simply looking for
$\tnp\ne 0$, this distinction is unimportant.)

Of course, theory will not, in general, predict a specific value for
$r$, but rather give a range. And in fact, theoretical estimates of
$r$, assuming no new physics, exist in the literature. Fleischer and
Mannel \cite{FM} quote the range
\beq
0.07 \leq r \leq 0.23 ~.
\label{fm}
\eeq
In view of the fact that Ref.~\cite{FM} does not include the $u$- and
$c$-quark contributions to the $b\to d$ penguin amplitudes, this range
must be expanded. We therefore take what we call the ``acceptable
range of $r$'' to be
\beq
0.05 \leq r \leq 0.5 ~.
\label{rbound}
\eeq

We should remark here that recent CLEO data \cite{CLEO} finds that the
branching ratios for $B\to K\pi$, which are dominated by $b\to s$
penguin amplitudes, are larger than expected. This suggests that the
$b\to d$ penguin amplitude may also be larger than expected. In
addition, CLEO finds that the branching ratio for $\bd \to \pi^+\pi^-$
is $4 \times 10^{-6}$, smaller than expected. Taken together, the data
suggest that the $P/T$ ratio may be quite a bit larger than the range
shown in Eq.~(\ref{rbound}), and that $P$ and $T$ interfere
destructively to reduce the $\bd\to\pi^+\pi^-$ branching
ratio\footnote{We note, however, that this naive picture is unlikely
  to be the full story. This explanation \cite{Kpidata} of the
  measured branching ratios requires $\cos\gamma < 0$, which is
  disfavoured by the SM \cite{AliLon}. Furthermore, the large
  branching ratio for $\bd\to K^0\pi^0$ \cite{CLEO} cannot be
  explained within this picture. It seems likely that more complicated
  effects, such as final-state interactions or inelastic scattering,
  are coming into play \cite{AtSoni,Zhouetal}.}. Nevertheless, since
the point of the paper is to explore the possibilities for finding new
physics in $B\to\pi\pi$, we will continue to use the range given in
Eq.~(\ref{rbound}). It may well be that, by the time measurements of
$B\to\pi\pi$ decays are done, the theoretical range for $r$ will have
changed. However, the techniques described in this paper for finding
new physics will still hold, since they do not depend on the exact
values chosen for the lower and upper bounds on $r$.

If, for a certain set of measurements, the value of $r$ obtained
assuming $\tnp=0$ is outside the range in Eq.~(\ref{rbound}), this
implies that new physics is present. One can then estimate the value
of $\tnp$ for which $r$ is lowered to the acceptable range. Of course,
this may not be feasible for all possible cases, and one may conclude
then that the large $r$ is due to new physics that does not contribute
simply to the phase of the penguin diagram, but also alters its
magnitude substantially.

\subsection{New Physics: Potential Difficulties}

In the previous subsection, we showed that, given measurements of
$\alpha$, $\adirpm$ and $\alphaeff$, along with a theoretical estimate
of the ratio of the penguin and tree amplitudes, one can extract
$\tnp$. If $\tnp$ is found to be nonzero, this will establish the
presence of new physics. In practice, however, the situation not quite
so simple.

First, as noted earlier, there are two ways that information about
$\alpha$ can be obtained: either through independent measurements
outside the $B\to\pi\pi$ system, or via an isospin analysis. One
problem with the isospin method is that it may not be so easy to make
all the measurements necessary to carry out the analysis. In
particular, the decays $\bd/\bdbar \to \pi^0\pi^0$ may be quite
challenging. Given that $B(\bd \to \pi^+ \pi^-)$ has been measured to
be $4 \times 10^{-6}$ \cite{CLEO}, this suggests that the branching
ratio for $\bd \to \pi^0 \pi^0$ is even smaller, perhaps considerably
so. In addition, the efficiency for the detection of the two $\pi^0$
mesons in the final state may not be very high. It is therefore
conceivable that it will not be possible to carry out a full isospin
analysis, at least at first-generation $B$-factories. (On the other
hand, if the $b\to d$ penguin is indeed large, as suggested by the
latest CLEO data \cite{CLEO}, then $\bd/\bdbar \to \pi^0\pi^0$ may
well be dominated by its penguin contributions, leading to a large
branching ratio. Indeed, recent analyses \cite{Zhouetal} of the CLEO
data, which includes inelastic rescattering effects, predict $B(\bd
\to \pi^0 \pi^0) \sim 5 \times 10^{-6}$. This branching ratio is about
an order of magnitude larger than earlier estimates based on
factorization. Thus, at this point in time, it is not clear how
difficult it will be to perform an isospin analysis of $B\to\pi\pi$
decays.)

Second, there are serious complications due to discrete ambiguities,
and this holds regardless of whether or not an isospin analysis is
done. Suppose, first, that one can perform the isospin analysis. In
this case, using isospin relations, the angle $\alpha$ is determined
by $2\alpha=2\alphaeff+\bar{\Phi}-\Phi$, where $\Phi$ and ${\bar\Phi}$
are obtained from Eq.~(\ref{cosdefs}). Since only $\cos\Phi$ and
$\cos\bar{\Phi}$ are known, there is a twofold ambiguity in each of
$\Phi$ and $\bar{\Phi}$, i.e.\ $\pm\Phi$ as well as $\pm\bar{\Phi}$
are allowed in the equation for $\alpha$. In addition, since it is the
quantity $\sineff$ which is measured, $2\alphaeff$ is also determined
up to a twofold ambiguity. Hence, $2\alpha$ is obtained with an
eightfold ambiguity.

The ratio $r^2$ itself has a fourfold ambiguity [see Eq.~(\ref{r})]:
the quantity $\cos(2 \alpha - 2 \alphaeff)$ takes two values, as does
$2\alphaeff$.  In general, then, we will find four distinct possible
values of $r^2$ for the same set of observables. This may make it
difficult to determine if new physics is present: if only one of the
four values of $r^2$ at $\tnp = 0$ lies within the acceptable range,
then the measurements may be consistent with the SM. One cannot
unequivocally conclude that there is new physics (though there might
be).

If the isospin analysis cannot be performed, then the CP phase
$2\alpha$ cannot be extracted cleanly from measurements of $\bd(t) \to
\pi^+ \pi^-$. In such a case, in order to use Eq.~(\ref{r}), we will
need to obtain knowledge of $2\alpha$ from other
measurements\footnote{Note that the independent knowledge of $\alpha$
  does not resolve the CKM ambiguity \cite{LSS}. The determination of
  $\alpha$ in the isospin analysis decouples from the solutions of the
  other parameters, and hence $\tnp$ still cannot be determined
  without fixing one of the theoretical parameters.}. This can be done
in several ways. For example, if the CP angles $\beta$ and $\gamma$
are extracted via $\bd(t) \to J/\Psi \ks$ and $B^\pm \to D K^\pm$,
respectively, this will give us information about $2\alpha$ due to the
unitarity triangle condition $2\alpha + 2\beta + 2\gamma = 0~~({\rm
  mod}~2\pi)$.  However, as will be explained in the next section,
this only determines $2\alpha$ up to a fourfold ambiguity, which,
along with the twofold ambiguity in $2\alphaeff$, still leaves an
eightfold ambiguity in $r^2$. A more promising source of information
is a $B\to \rho \pi$ Dalitz plot analysis \cite{Dalitz}. With this
method, one can obtain $2\alpha$ with no discrete ambiguity. In this
case, one is left with a twofold ambiguity in $r^2$.

Ideally, we will be able to perform a complete isospin $B\to \pi\pi$
analysis {\it and} have unambiguous knowledge of $2\alpha$ from $B\to
\rho\pi$. In this case, we will obtain a single value of $r^2$, which
will allow us to test unambiguously for the presence of new physics.

Unfortunately, in the real world we will probably have to deal with
one of the scenarios which gives $r^2$ with some number of discrete
ambiguities. In the next two sections, we will analyze all of these
scenarios. As we will see, even despite the presence of discrete
ambiguities, and even if a full isospin analysis cannot be performed,
there is still a significant region of parameter space where the
presence of new physics can be clearly established.

\section{Only $\bd(t) \to \pi^+ \pi^-$ is Measured}

We first suppose that only $\bd(t) \to \pi^+ \pi^-$ has been measured.
In this case, we will not have clean knowledge of the CP phase
$\alpha$. In order to use Eq.~(\ref{r}) to search for new physics, it
will then be necessary to obtain knowledge of $\alpha$ from
independent measurements. One possibility is to use the fact that,
even in the presence of new physics, the three angles $\alpha$,
$\beta$ and $\gamma$ still correspond to the interior angles of a
triangle \cite{nirsilv}. That is, we have $2\alpha + 2\beta + 2\gamma
= 0~~({\rm mod}~2\pi)$. Thus, measurements of $2\beta$ and $2\gamma$
will indirectly give us information about $2\alpha$, even if new
physics is present.

When one probes the CP phase $\beta$ via $\bd(t) \to J/\Psi \ks$, the
function one extracts is $\sin 2\beta$. This then determines $2\beta$
up to a twofold ambiguity. Similarly, the measurement of CP violation
in $B^\pm \to D K^\pm$ gives $\sin^2 \gamma$ (or equivalently $\cos
2\gamma$) which also yields $2\gamma$ up to a twofold ambiguity. Using
the triangle condition, these two measurements therefore determine
$2\alpha$ with a fourfold ambiguity. Since the measurement of $\bd(t)
\to \pi^+ \pi^-$ allows one to extract $\sin 2\alphaeff$, which
determines $2\alphaeff$ up to a twofold discrete ambiguity, in total
there is an eightfold ambiguity in the determination of $r^2$. With
such a large number of possible $r^2$ solutions, it is very likely
that at least one of them will lie within the acceptable $r^2$ region
[Eq.~(\ref{rbound})], in which case one cannot be sure that new physics
is present.

The situation can be improved in a variety of ways. There are methods
which use indirect, mixing-induced CP violation to extract functions
of $\beta$ and $\gamma$ other than $\sin 2\beta$ and $\sin
2\gamma$. This additional knowledge will remove the discrete ambiguity
in $2\beta$ and/or $2\gamma$. For example, a Dalitz-plot analysis of
the decay $\bd(t) \to D^+ D^- \ks$ allows one to extract the function
$\cos 2\beta$ \cite{Charlesetal}. This function can also be obtained
through a study of $\bd \to \Psi + K \to \Psi + (\pi^- \ell^+ \nu)$,
known as ``cascade mixing'' \cite{cascade}. Knowledge of both $\sin
2\beta$ and $\cos 2\beta$ determines $2\beta$ without
ambiguity. Similarly, $\sin 2\gamma$ can be obtained from $\bs(t)\to
D_s^\pm K^\mp$ if the width difference between the two $B_s$ mass
eigenstates is measurable \cite{IsiBs}, and this additional knowledge
removes the ambiguity in $2\gamma$. Finally, a Dalitz-plot analysis of
$\bd(t) \to D^\pm \pi^\mp \ks$ can be used to obtain the phase
$2(2\beta + \gamma)$ without ambiguity \cite{Charlesetal}, and this
knowledge will reduce the discrete ambiguity in both $2\beta$ and
$2\gamma$. Depending on which of these measurements are made, the
discrete ambiguity in $r^2$ can be reduced to a fourfold or even a
twofold ambiguity.

It is also possible to get at $2\alpha$ directly. If one performs a
Dalitz-plot analysis of $B\to \rho \pi$ decays, both $\sin 2\alpha$
and $\cos 2\alpha$ can be extracted \cite{Dalitz}. This then
determines $2\alpha$ with no ambiguity. In this case, one is left with
a twofold discrete ambiguity in $r^2$, due entirely to the discrete
ambiguity in $2\alphaeff$.

% This is Figure 1
\begin{figure}
%\vskip .050truein
%\centerline{\epsfysize 4.3truein \epsfbox {fig-1.eps}}
\centerline{\epsfxsize 6.0truein \epsfbox {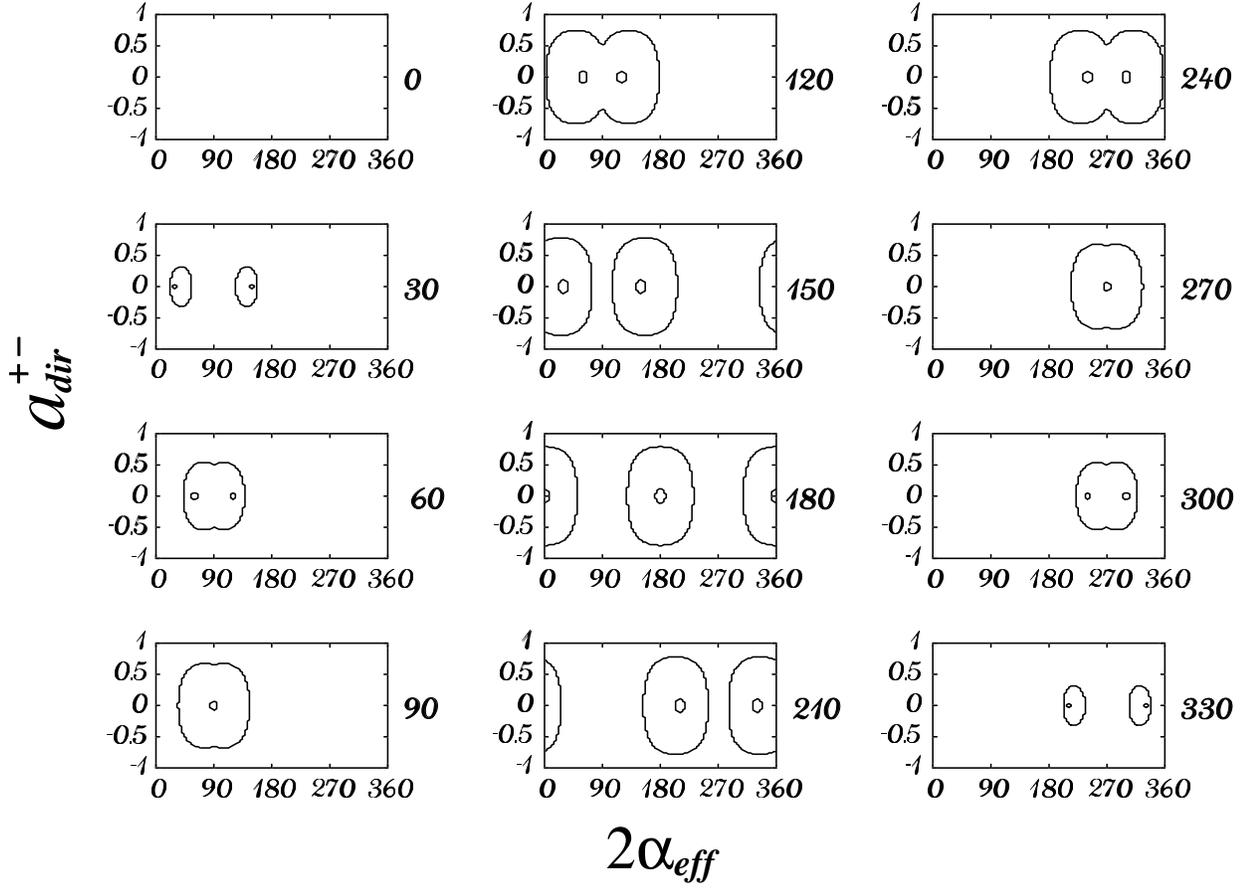}}
%\vskip -.1truein
\caption{The region in $2\alphaeff$--$\adirpm$ space which is
  consistent with the theoretical prediction for $|P/T|$
  [Eqs.~(\protect\ref{r}),(\protect\ref{rbound})], for various values
  of $2\alpha$. It is assumed that only $\bd(t)\to\pi^+\pi^-$ has been
  measured. In all figures, the $x$-axis is $2\alphaeff$ and the
  $y$-axis is $\adirpm$.  The value of $2\alpha$ used in a particular
  figure is given to the right of that figure.}
\label{Fig1}
\end{figure}

For all possible scenarios of this type, the prospects for discovering
new physics can be summarized in Fig.~\ref{Fig1}. We consider 12
specific values of $2\alpha$, varying between 0 and $2\pi$. For a
given value of $2\alpha$, we show the region in
$2\alphaeff$--$\adirpm$ space which is consistent with the SM. That
is, the region contains those values of $2\alphaeff$ and $\adirpm$ for
which the ratio $r$ satisfies the bound of Eq.~(\ref{rbound}). Note
that, for a given value of $2\alpha$, there are two allowed
$2\alphaeff$--$\adirpm$ regions. One of these regions is for
$2\alphaeff$, while the other corresponds to $\pi - 2\alphaeff$, which
reflects the fact that $2\alphaeff$ can only be measured up to a
twofold ambiguity.

Depending on which measurements have been made, $r^2$ will be
determined with an $N$-fold ambiguity ($N=2,4,8$). In a particular
scenario, in order to see whether the measurements indicate the
presence of new physics, one has to consider the $N$ values of the
pair $(2\alpha,2\alphaeff)$. If (at least) one of these sets of values
corresponds to a point in the appropriate plot which is consistent
with the SM, then one cannot conclude that new physics is present.
However, if all such values correspond to points in the plots which
lie outside the SM-allowed regions, then this is a clear signal of new
physics.

To give an example of how this works, suppose that the Dalitz-plot
$B\to\rho\pi$ analysis is performed, and it is found that $2\alpha =
180^\circ$ (present data suggests that $\alpha \simeq 90^\circ$ is the
preferred SM value \cite{AliLon}). If the measurement of $\bd(t) \to
\pi^+\pi^-$ yields $\sin 2\alphaeff = 0.966$ (i.e.\ $2\alphaeff =
75^\circ$ or $105^\circ$), then, regardless of the value of $\adirpm$,
this indicates the presence of new physics. On the other hand, if one
finds $\sin 2\alphaeff = 0$, then new physics is implied only if
$|\adirpm| \gsim 0.8$. 

We therefore see that the measurement of $\bd(t) \to \pi^+\pi^-$, when
combined with independent knowledge of $2\alpha$, can reveal the
presence of new physics, given a reliable prediction of the ratio
$P/T$ in the SM.

However, discrete ambiguities can muddy the picture considerably.  For
example, if $r^2$ is only known up to an 8-fold ambiguity, then one
must essentially superimpose 4 plots of the type shown in
Fig.~\ref{Fig1}, in which case there are very few values of the pair
($2\alphaeff$,$\adirpm$) which point unequivocally to new physics. For
this reason it is important to be able to reduce the discrete
ambiguity in $r^2$ as much as possible.

\begin{figure}
%\vskip -1.0truein
\centerline{\epsfysize 2.2 truein \epsfbox {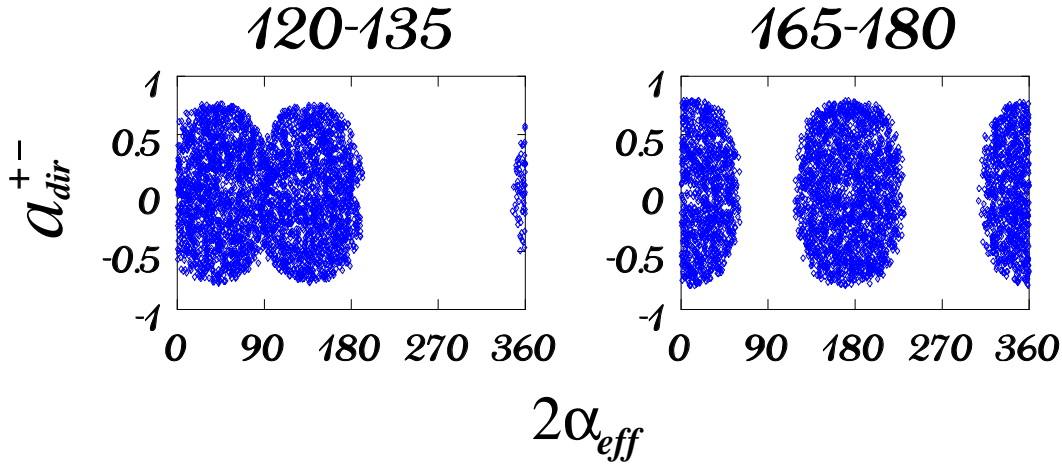}}
%\vskip .25truein
\caption{As in Fig.~\protect\ref{Fig1}, except that $2\alpha$ is 
  allowed to take a range of values. The range of values of $2\alpha$
  used in a particular figure is given above that figure.}
\label{Fig2}
\end{figure}

This point is made even sharper when one considers the fact that all
measurements will include experimental errors. In Fig.~\ref{Fig2}, we
assume that $2\alpha$ is known to be within a certain range
($120^\circ \le 2\alpha \le 135^\circ$ [left-hand figure of
Fig.~\ref{Fig2}] or $165^\circ \le 2\alpha \le 180^\circ$ [right-hand
figure of Fig.~\ref{Fig2}]). We then show the region in
$2\alphaeff$--$\adirpm$ space which is consistent with the SM. In this
case the allowed region is visibly larger than that presented in the
plots of Fig.~\ref{Fig1}. It is therefore clear that if, due to the
discrete ambiguity in $r^2$, one is forced to superimpose several such
figures, the prospects for detecting new physics will be considerably
reduced.

% This is Figure 2

Fortunately, the above analysis does not tell the whole story. Indeed,
this analysis is incomplete: it does not take into account the fact
that the isospin analysis must reproduce the independently-measured
value of $2\alpha$. Now, it is rather obvious that, if the decays
$B^+\to\pi^+\pi^0$ and $\bd/\bdbar \to \pi^0\pi^0$ can be measured,
and an isospin analysis performed, this additional constraint will
reduce the $2\alphaeff$--$\adirpm$ region which is consistent with the
SM. However it is also true that even if we have {\it no} information
about $B^+\to\pi^+\pi^0$ and $\bd/\bdbar \to \pi^0\pi^0$, the fact
that one must be able to reproduce $2\alpha$ using isospin is
sufficient to remove the twofold discrete ambiguity in $2\alphaeff$
which appears in all the plots of Figs.~\ref{Fig1} and \ref{Fig2}!
This remarkable result is discussed in the next section, along with an
examination of how actual measurements of, or limits on,
$B^+\to\pi^+\pi^0$ and $\bd/\bdbar \to \pi^0\pi^0$ can improve the
prospects for the detection of new physics.

\section{Beyond $\bd(t) \to \pi^+ \pi^-$}

In practice, it is likely that we will have more information about $B
\to \pi\pi$ decays than just the measurement of $\bd(t) \to \pi^+
\pi^-$. In the most optimistic scenario, the decays $B^+\to\pi^+\pi^0$
and $\bd/\bdbar \to \pi^0\pi^0$ will both be measured, which will
allow us to obtain the quantities $B^{+0}$, $B^{00}$ and $\adir00$. In
this case the full isospin analysis can be carried out, so that
$2\alpha$ can be determined. With this knowledge, one can then use
$r^2$ [Eq.~(\ref{r})] to search for new physics. (Note that, as
discussed in Sec.~2.1, in fact only the ratios of branching ratios
$B^{+0}/B^{+-}$ and $B^{00}/B^{+-}$ are needed to perform the isospin
analysis.)

However, there are problems with this procedure. First, there will be
errors associated with all measured quantities, which will lead to a
range of allowed values for $2\alpha$. Second, as discussed in
Sec.~2.3, the isospin analysis only determines $2\alpha$ and $r^2$ up
to an eightfold and fourfold ambiguity, respectively. As we saw in the
previous section, these two facts may make it difficult to
definitively establish the presence of new physics.

This situation can be improved if, in addition to the $B\to\pi\pi$
analysis, we have information about $2\alpha$ from other measurements.
In fact, this is quite likely: by the time $B^+ \to \pi^+ \pi^0$ and
$\bd/\bdbar \to \pi^0 \pi^0$ are measured, experiments which yield
independent information about $2\alpha$ will probably have been
performed. Various possibilities have been discussed in the previous
section. For example, the measurements of $\sin 2\beta$ and $\cos
2\gamma$, combined with the triangle relation $2\alpha + 2\beta +
2\gamma = 0~~({\rm mod}~2\pi)$, determine $2\alpha$ up to a fourfold
ambiguity. But this discrete ambiguity can be reduced by comparing
these four solutions with the eight obtained from the isospin
analysis. It is straightforward to see that, in general, there are
only two values of $2\alpha$ which are common to the four solutions
here and the eight solutions found in the isospin analysis. In
addition, for a given value of $2\alpha$, the value of $2\alphaeff$ is
fixed. {\it That is, the discrete ambiguity in $2\alphaeff$ which
  affected the analysis of Sec.~3 has been removed here}. The two
solutions are then
\beq
(2\alpha, 2\alphaeff) ~~,~~~~ (\pi - 2\alpha, \pi - 2\alphaeff) ~,
\eeq
and lead to a twofold ambiguity in $r^2$. Furthermore, if both $\sin
2\alpha$ and $\cos 2\alpha$ can be measured via a Dalitz-plot analysis
of $B\to \rho \pi$ decays \cite{Dalitz}, the remaining twofold
ambiguity will be lifted. In this case only the true $(2\alpha,
2\alphaeff)$ solution will remain, corresponding to a single value of
$r^2$. This is the key point: given an independently-determined value
of $2\alpha$, the isospin analysis removes the twofold discrete
ambiguity in $2\alphaeff$, and hence in $r^2$. As we will see below,
this is an important ingredient in searching for new physics.

Thus, by combining an isospin analysis with independent knowledge of
$2\alpha$, one can reduce the discrete ambiguity in $r^2$, thereby
improving the prospects for discovering new physics. In this section,
we assume that such independent knowledge of $2\alpha$ will in fact be
available.

The prescription to search for new physics then proceeds as follows.
For a given set of $B^{+0}/B^{+-}$, $B^{00}/B^{+-}$ and $\adir00$
measurements, we can calculate which values of $\adirpm$ and
$2\alphaeff$ produce values of $\alpha$ which lie within the measured
range. One can then check further to see which of these values of
$2\alphaeff$ and $\adirpm$ also give $r^2$ within the allowed
theoretical range [Eq.~(\ref{rbound})]. If the measured values of
$\adirpm$ and $2\alphaeff$ do not satisfy these two conditions, then
this is evidence for new physics.

Of course, in practice the measurements of $B^{+0}/B^{+-}$,
$B^{00}/B^{+-}$ and $\adir00$ will have errors associated with them.
In this case one can still use the above procedure, except that one
must scan over the allowed ranges for these quantities.

In the above discussion, we have assumed that the quantities
$B^{+0}/B^{+-}$, $B^{00}/B^{+-}$ and $\adir00$ have been actually
measured. However this may not turn out to be the case: since
$B^+\to\pi^+\pi^0$ and $\bd/\bdbar \to \pi^0\pi^0$ may be difficult to
measure, we may only have limits on these quantities. Fortunately,
assuming that independent information about $2\alpha$ will be
available, the above prescription can be carried out even in this
scenario. All that changes is that the allowed ranges for
$B^{+0}/B^{+-}$, $B^{00}/B^{+-}$ and $\adir00$ are (presumably) larger
than in the case where they are measured (with errors).

To illustrate how this all works, we consider a variety of
hypothetical experimental measurements. First, we take $2\alpha$,
assumed to have been obtained from measurements outside the
$B\to\pi\pi$ system, to lie within a given domain. (This corresponds
roughly to including an experimental error.) We consider two such
domains: (i) $165^\circ \le 2\alpha \le 180^\circ$ and (ii) $120^\circ
\le 2\alpha \le 135^\circ$.

Second, we assume that $\adir00$, $\B00/B^{+-}$ and $\Bp0/B^{+-}$ each
lie in a specified range.  For these allowed ranges, we consider five
distinct cases, shown in Table \ref{casetable}. In Case A, it is
assumed that $\bd/\bdbar \to \pi^0 \pi^0$ and $B^+ \to \pi^+ \pi^0$
have not been measured at all, so that we have no knowledge of
$\adir00$, $\B00/B^{+-}$ and $\Bp0/B^{+-}$. In Case B, the assumptions
are that (i) $\bd/\bdbar \to \pi^0\pi^0$ is not well-measured, so that
we have no knowledge of $\adir00$, and only an upper limit on
$\B00/B^{+-}$, and (ii) we have good knowledge of $B^{+0}/B^{+-}$ from
the measurement of $B^+\to\pi^+\pi^0$. In Case C, it is assumed that
(i) $\bd/\bdbar \to \pi^0\pi^0$ is well-measured, so that we have
rather precise knowledge of $\adir00$ and $\B00/B^{+-}$, but (ii) we
have only an upper limit on $B^{+0}/B^{+-}$. Finally, in Cases D and
E, all quantities are assumed to be known; only the range for
$B^{+0}/B^{+-}$ differs between the two cases.

\begin{table}
\hfil
\vbox{\offinterlineskip
\halign{&\vrule#&
 \strut\quad#\hfil\quad\cr
\noalign{\hrule}
height2pt&\omit&&\omit&&\omit&&\omit&\cr
& \omit && $\adir00$ && $B^{00}/B^{+-}$ && $B^{+0}/B^{+-}$ & \cr
height2pt&\omit&&\omit&&\omit&&\omit&\cr
\noalign{\hrule}
height2pt&\omit&&\omit&&\omit&&\omit&\cr
& Case A && $-1$ -- 1 && any value && any value & \cr
& Case B && $-1$ -- 1 && 0 -- 0.1 && 0.8 -- 0.9 & \cr
& Case C && 0.5 -- 0.7 && 0.7 -- 0.8 &&  0 -- 0.5 & \cr
& Case D && 0.6 -- 1 && 0.2 -- 0.4 && 0.6 -- 0.7 & \cr
& Case E && 0.6 -- 1 && 0.2 -- 0.4 && 0.2 -- 0.3 & \cr
height2pt&\omit&&\omit&&\omit&&\omit&\cr
\noalign{\hrule}}}
\caption{The assumed ranges for $\adir00$, $B^{00}/B^{+-}$ and 
$B^{+0}/B^{+-}$ for five (hypothetical) sets of experimental 
measurements.}
\label{casetable}
\end{table}

We are now in a position to apply the above prescription to search for
new physics. For a given range of $2\alpha$, and for a given case, we
use a random number generator to obtain values of $\adir00$,
$\B00/B^{+-}$ and $\Bp0/B^{+-}$ in the specified range, and $\adirpm$
and $2\alphaeff$ in the full allowed range. We generate $10^5$ sets of
values for these five parameters. For a given set, we then check to
see whether these parameters produce values for $2\alpha$ and $r^2$
which lie within their allowed ranges. In this way we map out the
region of $2\alphaeff$--$\adirpm$ space which is consistent with the
SM.

\begin{figure}
%\vskip -.25truein
\centerline{\epsfysize 6.5 truein \epsfbox {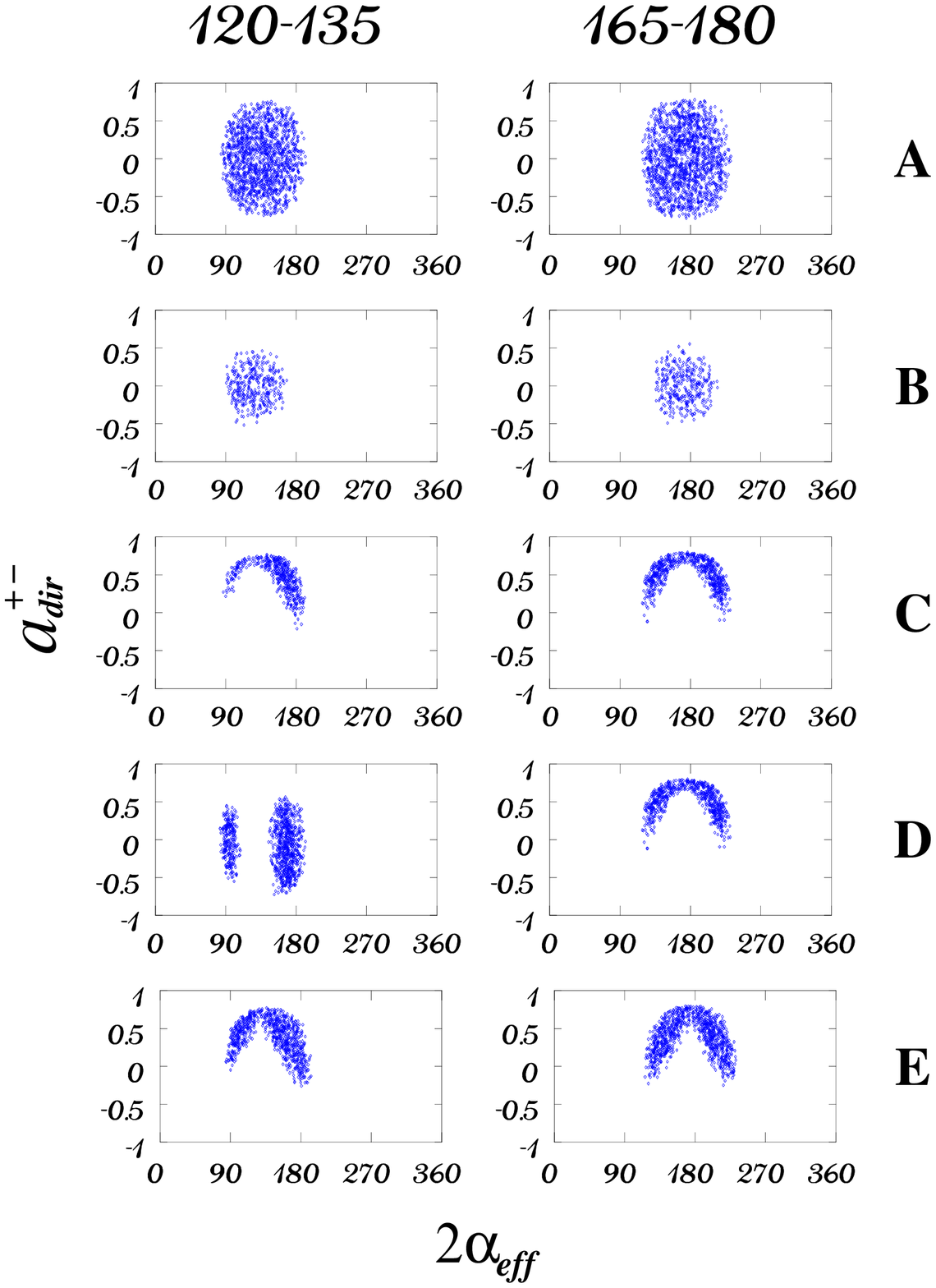}}
%\vskip -.25truein
\caption{The region in $2\alphaeff$--$\adirpm$ space which is
  consistent with the theoretical prediction for $|P/T|$
  [Eqs.~(\protect\ref{r}),(\protect\ref{rbound})]. In addition to the
  measurement of $\bd(t)\to\pi^+\pi^-$, it is assumed that information
  about $B^+\to\pi^+\pi^0$ and $\bd/\bdbar \to \pi^0\pi^0$ is
  available. For this latter information, the five scenarios of Table
  \ref{casetable} are considered from top (Case A) to bottom (Case E).
  In all cases, $2\alpha$ is allowed to take a range of values, given
  above each of the two columns of figures. In all figures, the
  $x$-axis is $2\alphaeff$ and the $y$-axis is $\adirpm$.}
\label{Fig3}
\end{figure}

The results are shown in Fig.~\ref{Fig3}. In all cases, by comparing
the SM-allowed $2\alphaeff$--$\adirpm$ region with that shown in
Fig.~\ref{Fig2}, one can see the extent to which the prospects for
detecting new physics are improved through considerations of isospin.

% This is Figure 3

Consider first Case A. Here, as was the case in Sec.~3, it is assumed
that we have no knowledge at all of $\adir00$, $\B00/B^{+-}$ and
$\Bp0/B^{+-}$. However, the difference here is that, despite this lack
of knowledge, we nevertheless require that an isospin analysis yield a
value of $2\alpha$ which lies in the assumed range. By comparing
Figs.~\ref{Fig2} and \ref{Fig3} for this case, one sees that this
condition is sufficient to remove one of the two solutions in
Fig.~\ref{Fig2}. In other words, for that solution, there are no
values of $\adir00$, $\B00/B^{+-}$ and $\Bp0/B^{+-}$ which will
simultaneously give $2\alpha$ and $r^2$ in their respective allowed
ranges. The removal of one solution will always occur as long as the
experimental range of $2\alpha$ is sufficiently restricted so as not
to include both $2\alpha$ and $\pi - 2\alpha$ values. This
demonstrates the power of the isospin analysis: even if $\bd/\bdbar
\to \pi^0 \pi^0$ and $B^+ \to \pi^+ \pi^0$ cannot be measured, the
isospin symmetry is able to remove the discrete ambiguity in
$2\alphaeff$ which appears in all the plots of Figs.~\ref{Fig1} and
\ref{Fig2}.

We now turn to Case B, in which it is assumed that the ranges for the
branching ratios $\B00/B^{+-}$ and $\Bp0/B^{+-}$ are reasonably well
known (though there is still only an upper limit on $\B00/B^{+-}$).
In this case, even though we still have no knowledge of $\adir00$,
Fig.~\ref{Fig3} shows that there is nevertheless a marked reduction in
the allowed $2\alphaeff$--$\adirpm$ region. Compared to the case where
only $\bd(t) \to \pi^+ \pi^-$ has been measured (Case A), the
$2\alphaeff$--$\adirpm$ region consistent with the SM has been reduced
by about a factor of two.

One can do even better if all of the three quantities $\adir00$,
$\B00/B^{+-}$ and $\Bp0/B^{+-}$ are measured reasonably well.
Depending on the measured values of these quantities, the allowed
$2\alphaeff$--$\adirpm$ region can be reduced even further, as the
plots for Cases C, D and E show.

Of course, in order to compute the full allowed
$2\alphaeff$--$\adirpm$ region, one will have to superimpose a certain
number of plots of this type, depending on the size of the discrete
ambiguity in $r^2$. As always, if the measured values of $2\alphaeff$
and $\adirpm$ lie outside the allowed region, then this will indicate
the presence of new physics.

Obviously, it is very unlikely that any of the hypothetical cases
considered above will turn out to be the actual experimental
situation. Indeed, even the theoretical situation --- namely the
allowed range for $r^2$ --- may change by the time the measurements
are done. However, regardless of the experimental and theoretical
numbers, the analysis described here can be used to search for new
physics.

\section{Conclusions}

In the near future, measurements will be made which will permit us to
extract the CKM angles $\alpha$, $\beta$ and $\gamma$ from
CP-violating rate asymmetries in the $B$ system. Hopefully, these
measurements will reveal the presence of new physics.

There are a variety of methods to test for new physics in the $b\to s$
flavour-changing neutral current (FCNC). However, there is no way to
{\it cleanly} detect new physics in the $b\to d$ FCNC. In order to
search for new physics in $b\to d$ transitions, one always needs some
theoretical input. That is, one needs to make an assumption regarding
hadronic parameters.

We have applied this idea to $B\to\pi\pi$ decays. If the decay
$\bd(t)\to\pi^+\pi^-$ were dominated by a tree-level amplitude ($T$),
then the angle $\alpha$ could be obtained with no hadronic
uncertainties. Unfortunately, this decay also receives a contribution
from a penguin amplitude ($P$) which may be sizeable, spoiling the
clean extraction of $\alpha$. However, it is well known that, by also
measuring the decays $\bd/\bdbar \to \pi^0\pi^0$ and $B^+ \to \pi^+
\pi^0$, one can use isospin to remove the penguin pollution and hence
obtain a clean measurement of $\alpha$. In this paper, we have shown
that, by making an assumption about the relative size of $P$ and $T$,
this isospin analysis can also be used to test for the presence of new
physics.

In fact, it is not even necessary to measure the decays $\bd/\bdbar
\to \pi^0\pi^0$ and $B^+ \to \pi^+ \pi^0$. If independent information
about $\alpha$ is available, then, given a prediction for the allowed
range of $|P/T|$, the measurement of the decay $\bd(t) \to \pi^+
\pi^-$ is sufficient to test for the presence of new physics. Here the
principle obstacle is the presence of discrete ambiguities. In the
simplest scenario, $2\alpha$ will probably only be known up to a
fourfold ambiguity. In this case, the measurement of $\bd(t) \to \pi^+
\pi^-$ yields eight possible values for $|P/T|$. By performing an
isospin analysis (which can be done even though no information from
$\bd/\bdbar \to \pi^0\pi^0$ and $B^+ \to \pi^+ \pi^0$ decays is
available!), one can reduce this to four possible $|P/T|$ values.
Still, even if there is new physics, it is quite likely that one of
these values will lie in the range for $|P/T|$ allowed by the standard
model, thereby masking the presence of new physics. Thus, in order to
search for new physics using only $\bd(t) \to \pi^+ \pi^-$, it will be
important to make additional measurements to reduce the discrete
ambiguity in $2 \alpha$.

If it is possible to perform a full isospin analysis, then independent
knowledge of $\alpha$ is not needed -- the isospin analysis itself
yields $2\alpha$. However, here too the presence of discrete
ambiguities may make it difficult to say with certainty that new
physics is present. On the other hand, by the time the full isospin
analysis is done, it is quite likely that we {\it will} have
independent information about $\alpha$. By combining this information
with that obtained from the isospin analysis, one can reduce the
discrete ambiguities substantially, thereby greatly improving the
prospects for detecting new physics.

In summary, we see that there are a variety of scenarios to consider,
depending on which measurements have been done. However, in all cases,
the bottom line is the following: the analysis of $B\to\pi\pi$ decays,
combined with a theoretical prediction for the allowed range of
$|P/T|$, can be used to search for the presence of new physics in the
$b\to d$ FCNC.

\section*{\bf Acknowledgments}

N.S. and R.S. thank D.L. for the hospitality of the Universit\'e de
Montr\'eal, where this work was initiated. The work of D.L. was
financially supported by NSERC of Canada.

\end{document}